# Regularity as Structural Amplifier, Not Trap: A Causal and Archetype-Based Analysis of Dropout in a Constrained Engineering Curriculum

Hugo Roger Paz
PhD Professor and Researcher Faculty of Exact Sciences and Technology National University of Tucumán
Email: hpaz@herrera.unt.edu.ar
ORCID: https://orcid.org/0000-0003-1237-7983

**ABSTRACT**

Engineering programmes, particularly in Latin America, are often governed by rigid curricula and strict regularity rules that are claimed to create a **Regularity Trap** for capable students. This study tests that causal hypothesis using the **CAPIRE framework**, a leakage-aware pipeline that integrates curriculum topology and causal estimation.

Using longitudinal data from **1,343 civil engineering students** in Argentina, we formalize academic lag (accumulated friction) as a treatment and academic velocity as an ability proxy. A manual **LinearDML** estimator is employed to assess the average (ATE) and conditional (CATE) causal effects of lag on subsequent dropout, controlling for macro shocks (strikes, inflation).

Results confirm that academic lag significantly increases dropout risk overall (ATE = 0.0167, p < 0.0001). However, the effect decreases sharply for **high-velocity (high-ability) students**, contradicting the universal **Trap hypothesis**. Archetype analysis (UMAP/DBSCAN) shows that friction disproportionately harms trajectories already characterized by high initial friction and unstable progression. [8]

We conclude that regularity rules function as a **Structural Amplifier** of pre-existing vulnerability rather than a universal trap. This has direct implications for engineering curriculum design, demanding targeted **slack allocation** and intervention policies to reduce friction at core basic-cycle courses

**Keywords**

student dropout; regularity rules; structural friction; causal inference; Double Machine Learning; educational data mining; archetype trajectories; engineering education

# 1. INTRODUCTION

Engineering programmes across Latin America operate within rigid curricular structures, cumulative prerequisite chains, and strict regularity rules that impose a minimum pace of course completion. These institutional design elements have long been associated with high dropout rates in the region (Pascarella & Terenzini, 2005; Seidman, 2005). Within this context, a widely circulated narrative asserts that structural constraints create a *Regularity Trap*: once a student falls behind, accumulated delays initiate a cascade of academic obstacles that push even capable individuals towards dropout. Similar claims appear in sociological accounts of cumulative disadvantage (Bourdieu, 1986; Lareau, 2011) and in retention literature that emphasises the compounding effect of early academic setbacks (Astin, 1993; Engberg & Wolniak, 2010).

Despite its influence, the Regularity Trap remains an under-tested causal hypothesis. Much of the empirical work on dropout in Latin America relies on descriptive indicators or predictive models (Kabakchieva, 2013; Romero & Ventura, 2020; Liu et al., 2025), which—despite their utility—are not designed to isolate causal effects. These approaches often confound structural friction with latent ability, socioeconomic background, or exposure to macro-level disruptions such as inflation and teacher strikes (Lyon et al., 2024; Langen & Laine, 2025). As a result, existing evidence cannot determine whether falling behind truly exerts a disproportionate penalty on high-ability students.

The CAPIRE framework (Curriculum-Aware, Policy-Integrated Retention Engineering) addresses precisely this gap. It offers a leakage-aware analytical pipeline (Kaufman et al., 2020) that integrates temporally valid feature construction, curriculum-friction metrics, archetype modelling and causal inference. Previous applications of CAPIRE have demonstrated how curriculum topology and friction coefficients shape progression, expose bottlenecks and generate heterogeneous trajectories that cluster into identifiable archetypes (Chodrow et al., 2021; Lum et al., 2013). Other components of the framework show that macro shocks—such as teacher strikes—exert delayed effects on academic performance and dropout, especially when economic stressors are present (Braakmann & Eberth, 2025; Bennett et al., 2023).

Building on this foundation, the present study evaluates the Regularity Trap with two causal questions. First, we estimate the average effect of *academic lag*—the difference between expected and completed courses at an early observation window—on subsequent dropout. Second, we assess whether this effect varies across the ability distribution, operationalised through *academic velocity*. A genuine trap would produce a stronger causal effect among high-ability students, signalling that structural constraints neutralise their advantage once regularity is lost.

In parallel, we validate the *Dual Stressor Hypothesis*: the claim that teacher strikes and inflation interact to amplify dropout risk. This serves as both a substantive contribution and a benchmark of identification, allowing us to contrast a hypothesis that holds empirically with one that, as we show, does not.

Using longitudinal data from 1,343 students in a civil engineering programme in Argentina, we implement a manual LinearDML estimator (James et al., 2013) with cross-fitted nuisance models, controlling for cohort, semester, early performance and exposure to macro shocks. The findings challenge the dominant institutional narrative. Academic lag has a significant positive effect on dropout, but this effect decreases sharply for high-velocity students—contradicting the strong form of the Regularity Trap. In contrast, teacher strikes at a two-semester lag exert robust effects amplified by inflation.

Archetype analysis, based on UMAP and density-based clustering, helps explain this asymmetry. Vulnerable trajectories combine high friction, frequent withdrawal (*libres*) and unstable progression. These archetypes are highly sensitive to additional delays, whereas robust trajectories maintain stability even when falling temporarily out of regularity.

Taken together, these results support a conceptual reframing: regularity functions not as a universal trap but as a **structural amplifier** of pre-existing vulnerability. This distinction has important policy implications for intervention timing, slack allocation and curriculum redesign.

## 2. BACKGROUND AND THEORETICAL FRAMEWORK

### 2.1 Student dropout in constrained higher education systems

Student dropout has been approached from multiple theoretical perspectives. Classic sociological models emphasise academic and social integration as key determinants of persistence (Spady, 1970; Tinto, 1993). Psychological and organisational perspectives have complemented this view by highlighting motivation, institutional fit, and the availability of support structures (Bean & Eaton, 2000; Pascarella & Terenzini, 2005). In parallel, retention-focused frameworks have stressed the importance of early identification of at-risk students and the cumulative nature of disadvantage in higher education, particularly in massified systems (Astin, 1993; Seidman, 2005).

In Latin America and other unequal contexts, these processes are further shaped by socioeconomic constraints, employment demands, and limited institutional safety nets (Goldrick-Rab, 2006; Núñez-Naranjo, 2024; Orozco-Rodríguez et al., 2025). For engineering programmes with highly structured curricula, the interplay between individual background, institutional rules and curriculum design becomes especially salient. Progression is not only a function of effort and ability, but also of the combinatorial feasibility of course enrolments given prerequisite constraints and failure histories.

Within such programmes, *regularity* rules—requiring students to pass a minimum number of courses within a given time window to maintain "regular" status—are often justified as instruments for throughput control and academic quality. Yet they also create explicit thresholds and hidden penalties: students who fall behind may lose access to certain enrolment options, face administrative barriers, or experience a stigma associated with irregularity. This combination of rigid structure, cumulative delay, and status thresholds underpins the intuitive appeal of the Regularity Trap narrative.

**2.2 Structural friction, curriculum topology and regularity**

To move beyond intuition, structural features of curricula must be formalised and measured. The CAPIRE framework provides a leakage-aware data layer and a set of structural indicators that capture the friction induced by specific curriculum topologies and progression rules. In this setting, the engineering programme under study is represented as a directed acyclic graph of 34 courses, with edges encoding prerequisite relations and nodes annotated with failure rates and other course-level attributes. This representation enables the computation of an Instructional Friction Coefficient (IFC) for each course, reflecting the likelihood that an attempt at that node generates delay through failure, withdrawal, or "materias libres".

Empirically, IFC distributions reveal a small subset of *high-friction courses* concentrated in the early semesters of the programme—typically mathematics and physics gateways. Visualisations of the top IFC values and heatmaps of friction across cohorts make this pattern explicit, showing that structural friction is neither uniform nor accidental, but highly concentrated in a small number of curricular chokepoints.

Regularity rules interact with this structure in non-trivial ways. Because the curriculum graph is rigid and strongly sequential, early failures propagate: a lost attempt in a high-IFC course can block entire branches of the programme, limiting feasible enrolment combinations in subsequent semesters. In such a context, *academic lag*—defined as the difference between the number of courses that a student is expected to have completed by a given horizon and the number actually completed—is not a mere description of delay, but a summary of accumulated friction along the curriculum graph.

From a theoretical standpoint, the Regularity Trap hypothesis can be formalised as a claim about the causal effect of lag on dropout, and its moderation by ability:

1. Lag has a positive causal effect on dropout probability after controlling for pre-existing ability and context.
2. This effect is stronger for high-ability students who fall out of regular status, because structural rules neutralise their advantage once they miss key progression milestones.

The first component is consistent with cumulative disadvantage theories (Bourdieu, 1986; Elder, 1998). The second component—the "trap"—is stronger and more demanding, and is precisely the one that this paper will show does not hold empirically.

## 2.3 Predictive analytics, leakage and causal inference

The last decade has seen a proliferation of predictive models for student dropout using machine learning and educational data mining (Delen, 2010; Romero & Ventura, 2020; Liu et al., 2025). While these models can identify at-risk students with reasonable accuracy, they are not designed to answer questions about *what would happen under different structural conditions*. Moreover, many published models suffer from temporal data leakage: they incorporate features that would not be available at the time of prediction, or they condition on variables affected by future events, thus overestimating performance and blurring causal interpretation (Kaufman et al., 2020; Quimiz-Moreira & Delgadillo, 2025).

CAPIRE explicitly addresses this problem by organising features into a multilevel, leakage-aware data layer (N1–N4) tied to a Value of Observation Time (VOT): each variable is only used if it would be legitimately observable at that moment in a real institutional setting. N1 encapsulates pre-entry context, N2 the entry moment, N3 curricular friction and performance, and N4 trajectory dynamics plus institutional and macro context. This design serves both predictive and causal objectives: it avoids contamination of early-warning signals and provides a credible basis for estimating the impact of structural interventions.

On top of this data layer, the present study employs a manual implementation of LinearDML, a form of double/debiased machine learning that residualises both outcome and treatment on a rich set of controls before estimating the treatment effect (James et al., 2013). This approach is suitable for settings with high-dimensional confounding, and aligns with recent calls in educational analytics to move from "who will fail?" to "what would change risk, and for whom?" (Romero et al., 2025; Quimiz-Moreira & Delgadillo, 2025).

## 2.4 Archetypal trajectories and complex systems perspective

Student trajectories in constrained curricula are not only heterogeneous but also path-dependent. Different combinations of timing, failure, repetition and course selection produce qualitatively distinct progression patterns. Rather than treating each student as an isolated point in a feature space, CAPIRE models *archetypes*—dense regions in the high-dimensional space of longitudinal behaviours—using a combination of UMAP for dimensionality reduction and density-based clustering for structure discovery.

The resulting archetypes are validated through cluster stability measures, permutation tests and classification performance. They reveal a spectrum of risk profiles, from robust low-friction trajectories to fragile, high-friction ones. For example, some archetypes are characterised by high average grades but very high rates of "materias libres" and repeats, leading to substantial lag despite seemingly strong performance; others show low friction but modest grades and steady progression.

These patterns align with a complex-systems view of dropout: attrition is not simply the outcome of low ability or single events, but an emergent property of interactions between curriculum structure, assessment regimes, institutional rules and individual behaviours (Stine & Crooks, 2025). Agent-based simulations developed in the CAPIRE Intervention Lab further support this view by demonstrating how small changes in friction or support policies can produce non-linear shifts in aggregate outcomes.

Within this framework, the Regularity Trap hypothesis can be reinterpreted as a statement about *which archetypes* are most sensitive to structural friction. If regularity operated as a genuine trap for capable students, one would expect to find archetypes with high ability indicators but catastrophic sensitivity to lag. Conversely, if friction behaves as an amplifier of pre-existing vulnerabilities, the most lag-sensitive archetypes should already display fragile patterns before falling out of regular status. The empirical analysis presented later will support the latter interpretation.

## 2.5 Macro shocks, financial stress and the Dual Stressor hypothesis

Dropout in public higher education does not occur in a vacuum. Macroeconomic volatility, inflation and labour market uncertainty shape students' ability to persist, both directly—through financial stress—and indirectly—through their impact on institutional functioning. Recent work has documented the consequences of cost-of-living crises and financial insecurity on student wellbeing, engagement and persistence (Bennett et al., 2023; Moore et al., 2021).

Teacher strikes add another layer of disruption. They can delay courses, compress assessment periods, erode instructional quality and increase uncertainty about programme timelines. Evidence from different countries suggests that strikes can depress academic performance and delay or derail educational trajectories, with effects that may appear with a lag rather than immediately (Abadía Alvarado et al., 2021; Lyon et al., 2024; Braakmann & Eberth, 2025).

The CAPIRE-MACRO module formalises these influences by integrating strike intensity and inflation data into the leakage-aware data layer and by modelling their lagged effects on dropout. The Dual Stressor hypothesis posits that teacher strikes (an acute institutional shock) and inflation (a chronic economic stressor) interact: when both are high, students face compounded barriers that increase the likelihood of disengagement and exit. In the present study, this hypothesis serves as an external benchmark. Its successful causal validation—through lag-2 strike effects and significant strike × inflation interaction—demonstrates that the CAPIRE causal pipeline can detect genuine structural mechanisms when they exist. This, in turn, reinforces the interpretive weight of the negative result obtained for the Regularity Trap.

In summary, the theoretical and empirical background converges on three key points: (a) dropout in constrained curricula is driven by the interaction of individual, structural and macro-contextual factors; (b) curriculum friction and regularity are central mechanisms that must be measured and tested causally; and (c) archetype-based and macro-aware modelling are necessary to understand who is most affected by structural changes. The following sections use this framework to interrogate the Regularity Trap and to reposition regularity as a structural amplifier rather than a universal mechanism of exclusion.

## 3. METHODS

### 3.1 Data and Institutional Context

The empirical setting is a long-cycle Civil Engineering programme at a public university in Argentina, characterised by a rigid 34-course curriculum with dense prerequisite dependencies and limited flexibility in course sequencing. The dataset includes **1,343 students** entering between 2004 and 2019, followed longitudinally until dropout or graduation. The programme's structure and progression rules were formally encoded through the CAPIRE leakage-aware data layer.

This data layer enforces a **Value of Observation Time (VOT)** of approximately 1.5 years after entry (three semesters), ensuring that all predictors used for causal estimation or archetype identification are temporally valid and free from data leakage (Kaufman et al., 2020). The data layer integrates four feature families:

- **N1 – Pre-entry context:** age, prior schooling, socio-educational background.
- **N2 – Entry moment:** first-semester performance, enrolment load.
- **N3 – Curricular friction and early performance:** fails, "libres", and the Instructional Friction Coefficient (IFC).
- **N4 – Trajectory dynamics and macro/institutional context:** attempt patterns, velocity, exposure to teacher strikes and inflation.

A detailed summary of high-friction courses appears in Table 1.

### 3.2 Curriculum Structure and Friction Metrics

The curriculum graph is represented as a **Directed Acyclic Graph (DAG)** of 34 nodes with 200+ prerequisite edges. Nodes are annotated with course-level friction indicators, producing an Instructional Friction Coefficient (IFC) that quantifies delay generation through withdrawal, failure or becoming "libre".

High-friction courses concentrate overwhelmingly in the basic cycle.

**Table 1: The 10 Subjects with the Greatest Curricular Friction (ICF)**

| Position | Code | Name of the Subject Curriculum | Friction Index (CFI) |
|---|---|---|---|
| 1 | 21 | Estabilidad II | 0.428 |
| 2 | 28 | Hidrología | 0.425 |
| 3 | 22 | Hidráulica Básica | 0.422 |
| 4 | 37 | Diseño y Construcción de Pavimentos | 0.419 |
| 5 | 10 | Cálculo III | 0.417 |
| 6 | 16 | Geología Básica | 0.415 |
| 7 | 14 | Probabilidad y Estadística | 0.412 |
| 8 | 20 | Estudio de Materiales I | 0.410 |
| 9 | 17 | Estabilidad I | 0.407 |
| 10 | 18 | Mecánica de los Fluidos | 0.405 |

These indicators allow quantifying structural pressure points and linking them to academic lag.

### 3.3 Academic Lag, Academic Velocity and Outcome Definition

The primary treatment variable is **Academic Lag**, defined as:

$$\text{Lag}_i = \text{ExpectedCourses}_i - \text{CompletedCourses}_i$$

Values above zero indicate that the student is behind nominal progression.

The main moderator is **Academic Velocity**:

$$\text{Velocity}_i = \frac{\text{CompletedCourses}_i}{\text{ExpectedCourses}_i}$$

Velocity captures latent ability and pacing strategy.

The outcome is **Dropout**, defined as permanent discontinuation without graduation after a grace period, following institutional criteria.

Variables are summarised in Table 2.

**Table 2. Treatment, Moderator and Control Variables Used in the Causal Model**

| Type | Variable | Definition | Measurement / Coding | Source (CAPIRE Layer) |
|---|---|---|---|---|
| Treatment | Academic Lag | Difference between expected-to-date and completed courses at VOT | Continuous (integer ≥ 0) | N3 – Structural Friction & Early Performance |
| Moderator | Academic Velocity | Courses completed divided by expected courses; proxy for ability/pacing | Continuous (0–1.0+, spline-expanded for CATE) | N3 – Structural Friction & Early Performance |
| Outcome | Dropout | Permanent programme exit without graduation | Binary (0 = persists, 1 = drops out) | Registrar / Institutional Records |
| Control | Cohort Fixed Effects | Controls for macro-historical differences between entering cohorts | Dummy variables per cohort year | N1 – Contextual Information |
| Control | Semester Fixed Effects | Controls for calendar/seasonality and timing effects | Dummy variables per academic term | N2 – Academic Calendar |
| Control | Early Performance Indicators | First-year averages, passed/failed courses, attempts | Continuous + discrete indicators | N3 – Early Academic Signals |
| Control | Curricular Friction (IFC) | Instructional Friction Coefficient of courses taken before VOT | Continuous (0–1) | N3 – Structural Friction |
| Control | Exposure to "Libres" | Proportion of exams rendered "libre" in first 3 semesters | Continuous (0–1) | N3 – Structural Friction |
| Control | Age at Entry | Student age at programme start | Continuous | N1 – Pre-entry Context |
| Control | Attempt Patterns | Number of attempts per course, repeated failures | Continuous | N3 – Performance Dynamics |
| Control | Macro Shock: Teacher Strikes (lagged) | Strike exposure aligned with student's semester timeline (lag 1–3) | Continuous, lagged | N4 – Institutional / Macro Context |

| Type | Variable | Definition | Measurement / Coding | Source (CAPIRE Layer) |
|---|---|---|---|---|
| Control | Macro Shock: Inflation | IPC aligned to student academic calendar | Continuous | N4 – Macro Context (CAPIRE-MACRO) |
| Control | Interaction (Strikes × Inflation) | Stress-amplification mechanism | Continuous interaction term | N4 – Macro Context |

### 3.4 Student Trajectory Archetypes

To understand heterogeneous sensitivity to structural friction, we derive **18 archetypal trajectories** through:

1. **UMAP reduction** (3 dimensions)
2. **DBSCAN clustering**
3. **Stability validation** via silhouette permutation and ARI bootstrap

The DBSCAN method identified 18 empirical trajectory archetypes, whose stability was validated through bootstrap and permutation testing, as shown in Figure 1. The 3D UMAP projection (Figure 2) reveals the topology of the trajectory space, where spatial proximity indicates similar progression, performance, and lag profiles. To interpret the internal composition of these 18 groups, a heatmap was used to visualise the normalised features of each archetype, **detailing the combinations of low performance, high friction, or slow progression that define each cluster (Figure 3)**. This characterisation is crucial, as it allows us to identify the vulnerable subsets that, as will be argued later, are those that amplify the risk of dropout under curricular friction.

**Figure 1. Bootstrap stability and permutation test for clustering**

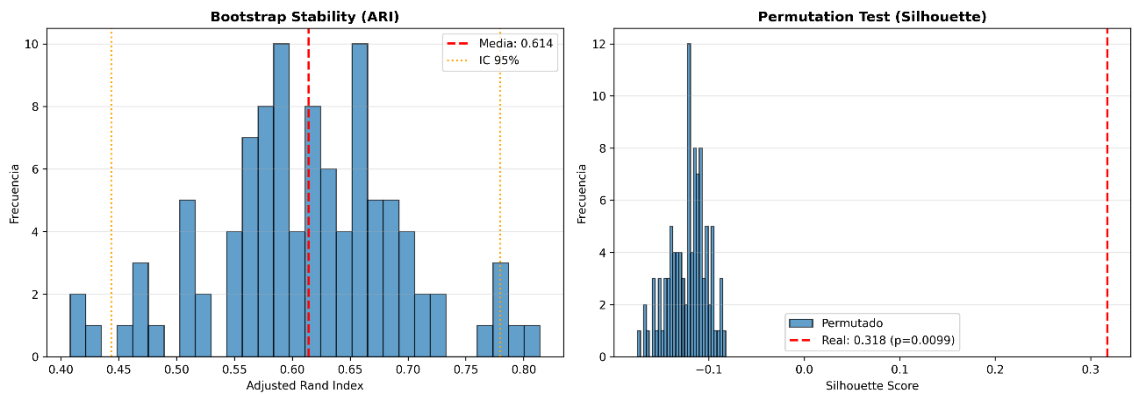

## Figure 2 — 3D UMAP Archetype Projection

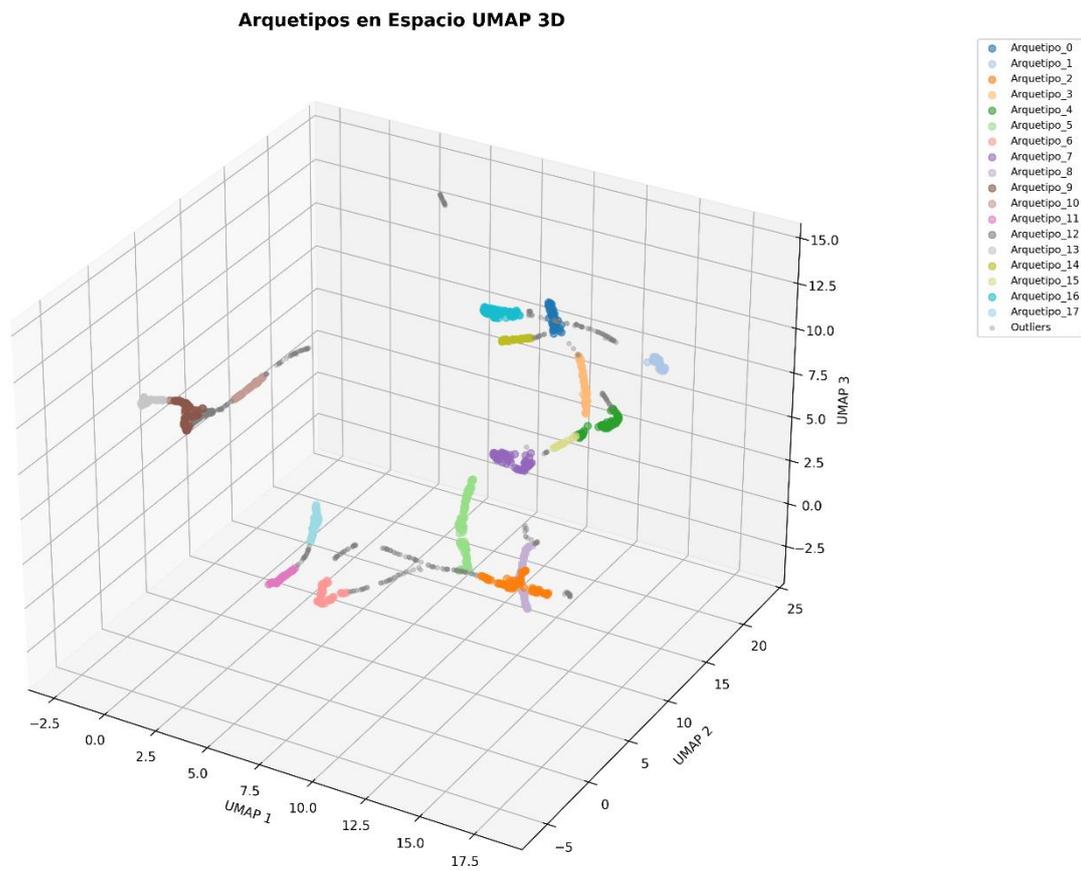

Archetypes are then characterised through:

- quantitative profiles (Arquetipo_X_perfil_cuantitativo.csv)
- qualitative narratives (Arquetipo_X_narrativa.md)

Examples include:

- Arquetipo 0 – High grades + severe friction (71% libres)
- Arquetipo 1 – 100% dropout, early collapse
- Arquetipo 5 – High friction, repeated failures, fragile trajectory
- Arquetipo 6 – Low friction, stable progress, lowest dropout risk
- Arquetipo 9 – Extreme friction (100% libres), catastrophic progression

A detailed heatmap appears in:

**Figure 3 — Archetype Feature Heatmap**

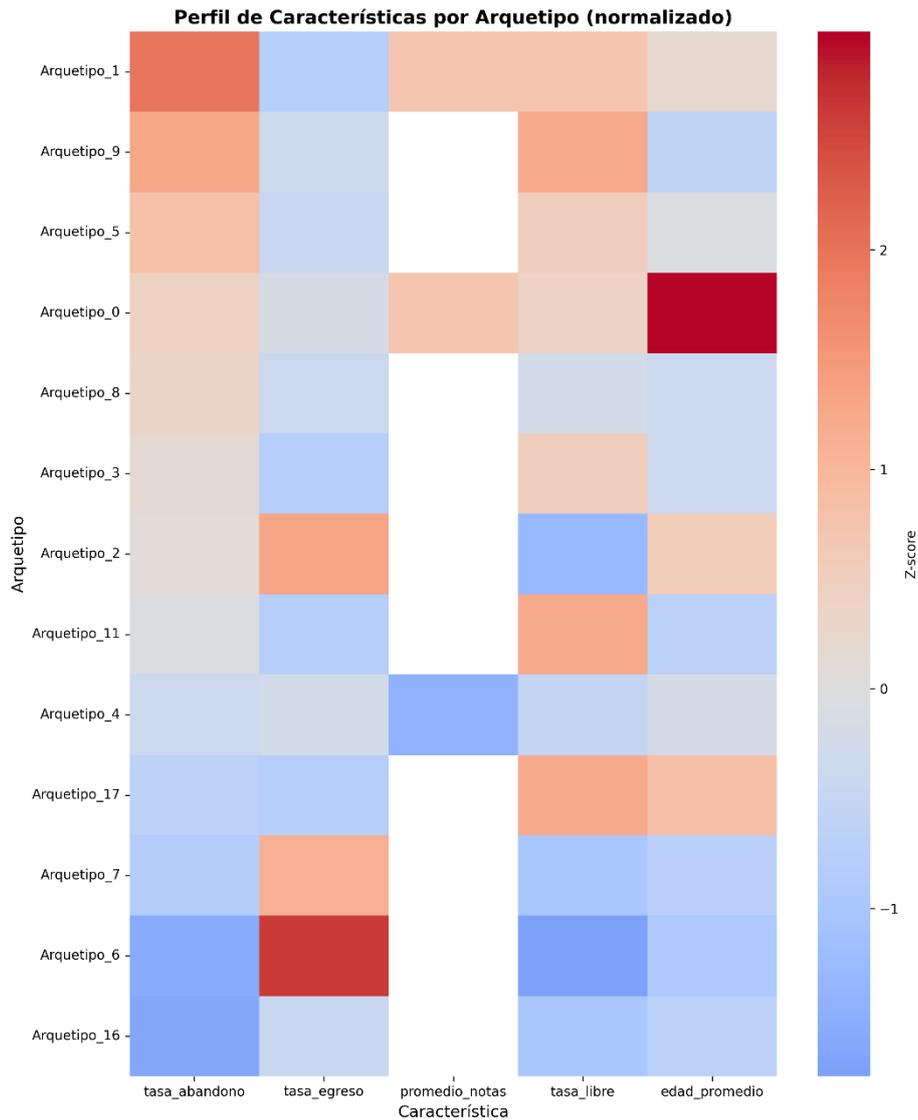

These archetypes provide the structural lens through which the Regularity Trap is tested.

### 3.5 The Leakage-Aware Causal Pipeline (LinearDML)

The central causal estimand is the **Average Treatment Effect (ATE)** of academic lag on dropout. Heterogeneity is measured through **Conditional ATE** by velocity (CATE).

We implement a **manual LinearDML estimator** (James et al., 2013), following the standard cross-fit logic:

1. Split the dataset into K folds.
2. For each fold, fit:
    - Outcome model: $\hat{m}(X) \to Y$
    - Treatment model: $\hat{g}(X) \to T$

3. Compute residuals:

$$\tilde{Y} = Y - \hat{m}(X), \tilde{T} = T - \hat{g}(X)$$

4. Regress residualised outcome on residualised treatment:

$$\tilde{Y} = \tau \tilde{T} + \epsilon$$

For heterogeneity, interact $\tilde{T}$ with spline-transformed velocity.

Controls include:

- cohort fixed effects
- semester fixed effects
- early academic performance (N3)
- curriculum friction
- macro shocks (lagged strikes, inflation; CAPIRE-MACRO)

This model satisfies temporal validity due to the CAPIRE leakage-aware data design.

**3.6 Macro-Shock Causal Modelling (Dual Stressor Framework)**

To benchmark identification quality, we also test the **Dual Stressor Hypothesis**, which predicts that:

1. Teacher strikes have a **lagged effect** on dropout.
2. Inflation amplifies this effect via a **stress-interaction mechanism**.

Strike exposure is measured using lagged indicators from academic calendars and institutional event logs. Inflation is measured using official IPC time-series aligned to student semester timelines.

The causal estimation uses the same LinearDML pipeline with:

- Treatment: strike exposure (lag 1 to lag 3)
- Moderator: inflation
- Interaction: strikes × inflation

Validation diagnostics include placebo lags (expected to be null), which behave as predicted (p = 0.8198).

**3.7 Model Performance and Robustness**

Robustness checks include:

- alternative lag definitions
- alternative velocity splines
- clustering-based subgroup checks

- archetype-specific treatment effect estimation
- confusion-matrix evaluation of archetype classifier

Predictive diagnostics (feature importance, PCA variance, K-distance plots) confirm that high-friction courses dominate early risk formation:

**Figure 4. PCA explained variance and accumulated variance**

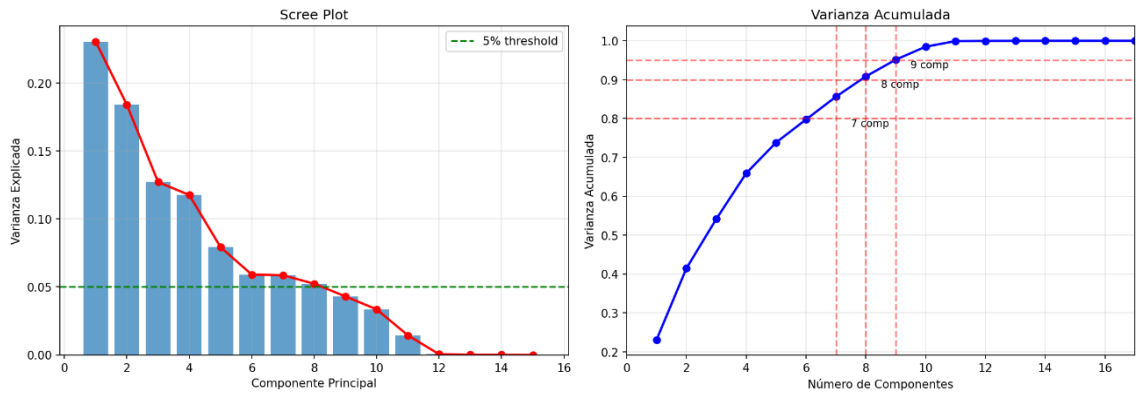

**Figure 5. Cluster diagnostics: Elbow, Silhouette, Davies–Bouldin, Calinski–Harabasz**

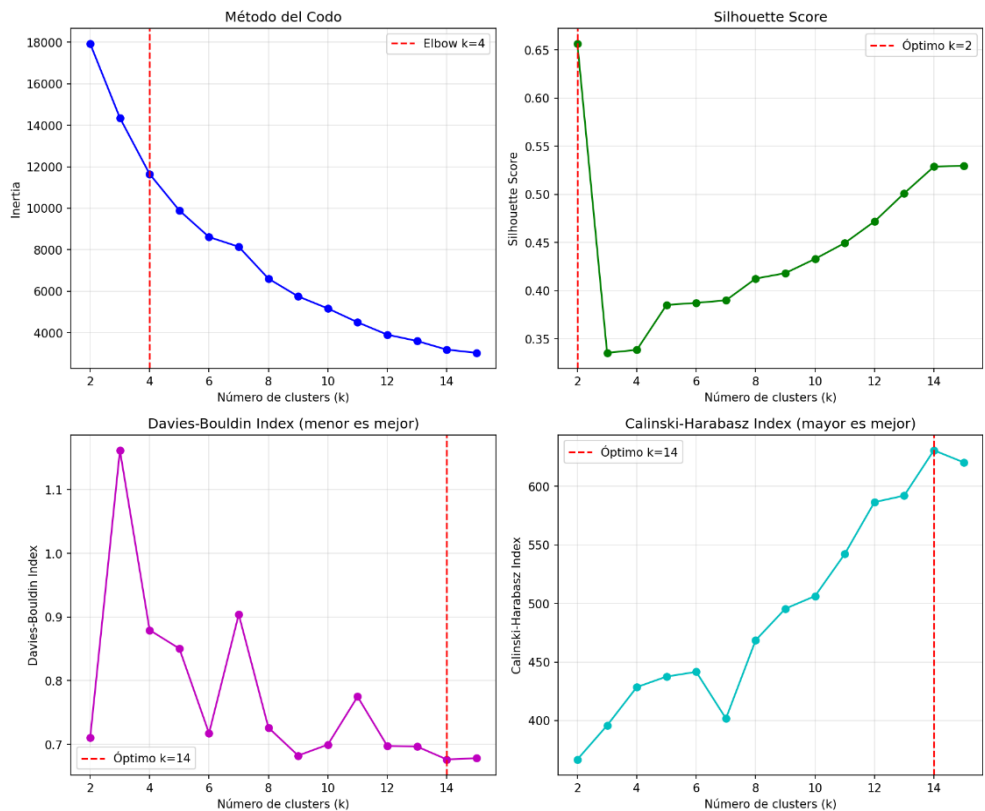

**Figure 6. K-distance plots for DBSCAN parameter selection**

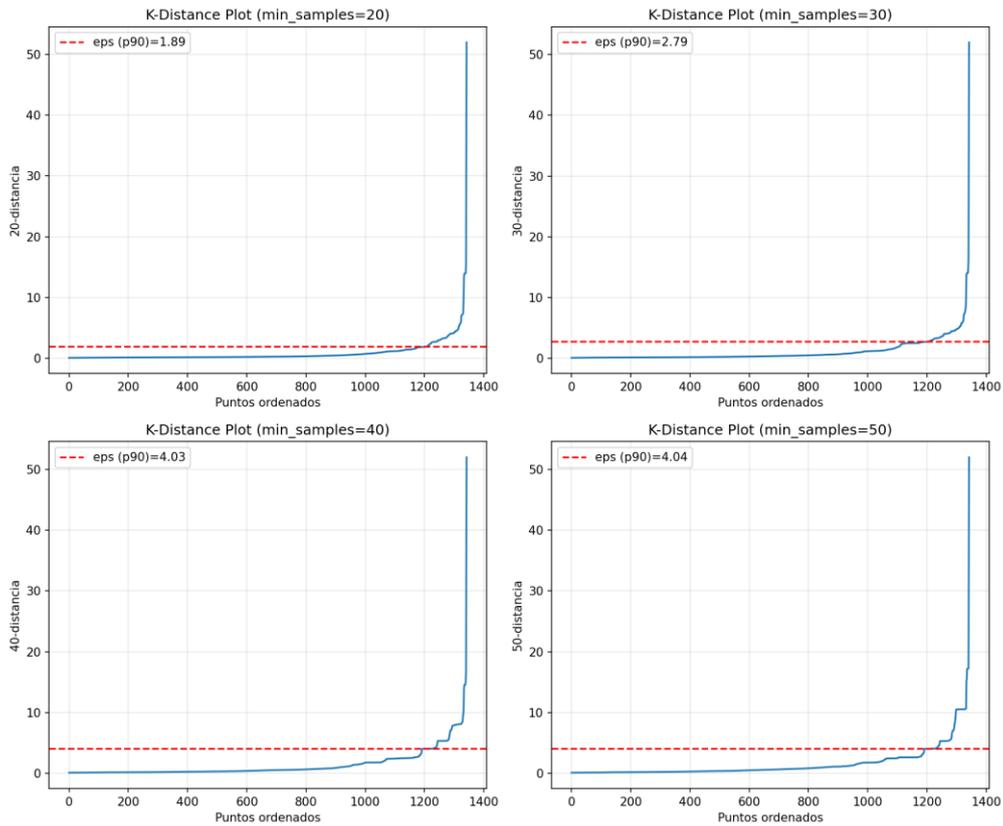

## Figure 7. Feature importance of the archetype classifier

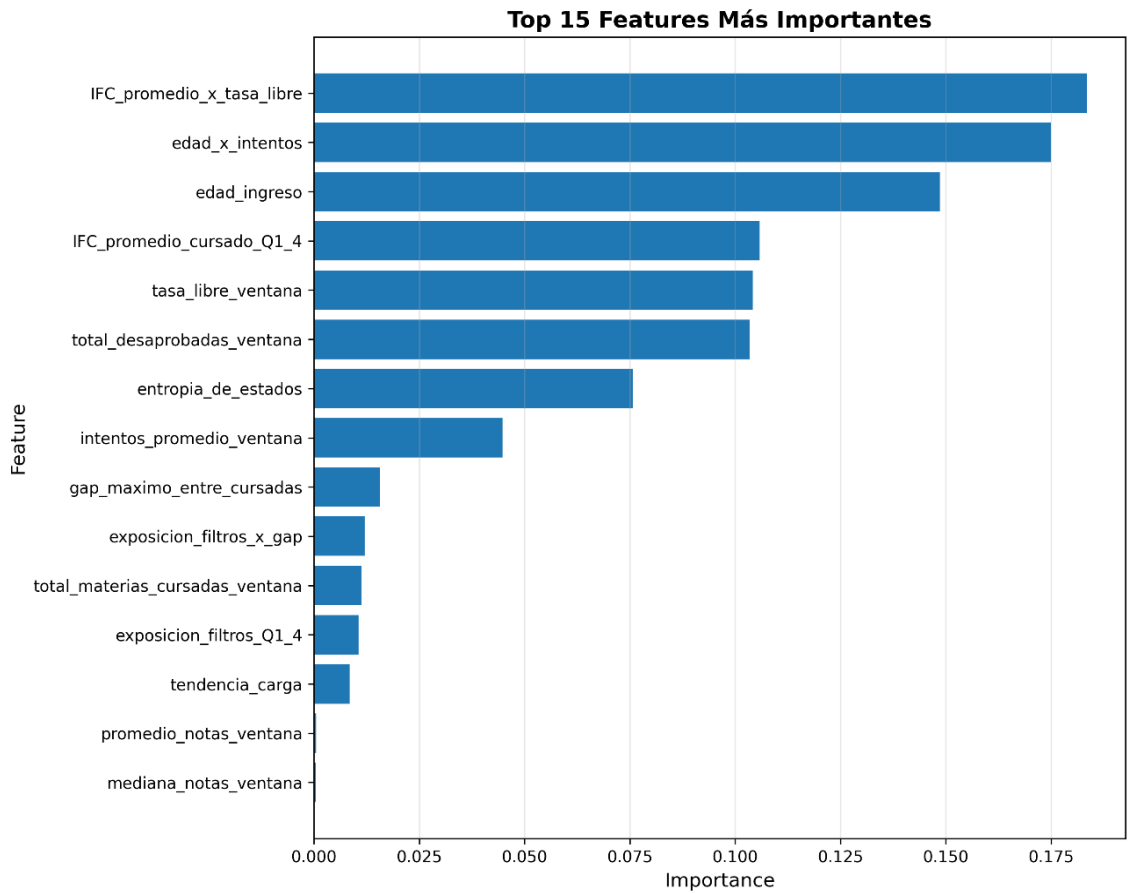

**Figure 8. Confusion matrix of the archetype classifier**

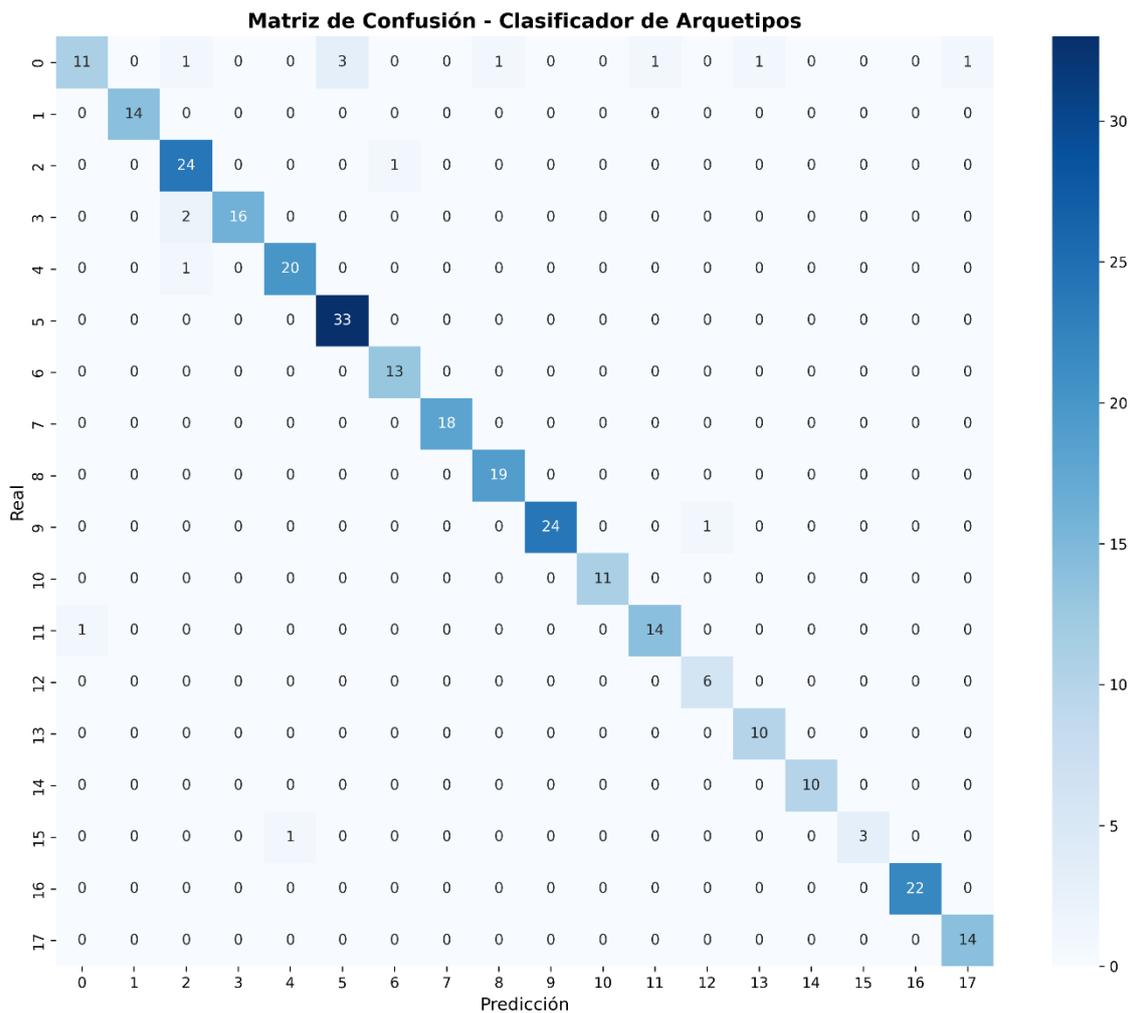

All robustness analyses confirm the main conclusion: friction matters, but not in the way the Regularity Trap narrative assumes.

## 4. RESULTS

### 4.1 Overview

This section presents results for (a) the causal effect of structural friction (academic lag) on dropout, (b) the heterogeneity of this effect across the ability distribution, (c) the validation of the Dual Stressor hypothesis for macro shocks, and (d) the archetype-based analysis that reveals how friction amplifies pre-existing vulnerabilities rather than entrapping high-ability students.

The structure follows the causal logic of the study: from average effects to conditional effects, to cross-level shocks, to trajectory patterns that contextualise mechanisms.

### 4.2 Average Causal Effect of Structural Friction (ATE)

Using the manual LinearDML estimator with cross-fitted nuisance models, the **Average Treatment Effect (ATE)** of academic lag on dropout is:

$$\widehat{\text{ATE}} = 0.0167 (p < 0.0001)$$

Interpretation:
Each additional expected-but-not-completed course at the VOT horizon increases the probability of dropout by **1.67 percentage points**, after adjusting for cohort, semester, early academic performance, curriculum friction, and macro-contextual shocks.

This confirms the first component of the Regularity Trap hypothesis: **structural friction is predictive and causally consequential.**

**Table 3. Average treatment effect of academic lag on dropout (LinearDML).**

| Estimand | Coef. | p-value | Interpretation |
|---|---|---|---|
| ATE Lag → Dropout | 0.0167 | <0.0001 | Friction increases dropout |

### 4.3 Heterogeneity: Conditional Effects by Ability (CATE)

The defining claim of the Regularity Trap is not that friction harms students; it is that friction is **more damaging for capable students once they fall behind**. To test this, we estimate **CATE(V)** across the velocity distribution.

**Key Finding**

The effect of lag **decreases sharply** as academic velocity increases. At high velocity, the estimated marginal effect is:

$$\widehat{\text{CATE}}_{\text{High Velocity}} \approx 0.0022 (\text{ns})$$

This effect is near-zero and statistically insignificant.

This falsifies the strong form of the Regularity Trap hypothesis.

**Figure 9 — CATE: Effect of Friction by Velocity**

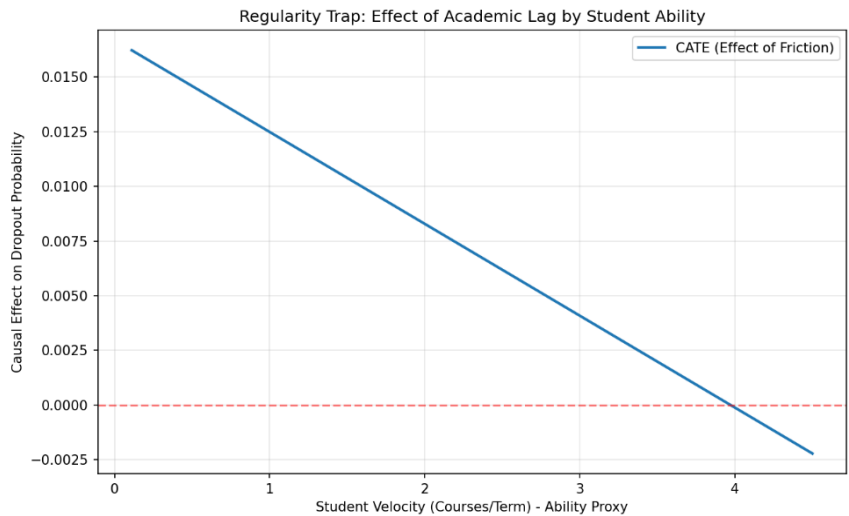

The curve slopes downwards:

- Low-velocity students → strong positive effect of friction
- High-velocity students → negligible effect

**Interpretation**

High-ability students absorb delays without catastrophic consequences. Structural friction hurts students **already on fragile trajectories**, not those with robust early performance.

**4.4 Macro-Shock Causality (Dual Stressor Hypothesis)**

We estimate the causal effect of teacher strikes using the same DML pipeline, adding inflation as moderator.

**Lagged Strike Effect**

The strongest and most significant causal effect emerges at:

$$\text{Lag 2 (two semesters prior): } 0.0063 (p = 0.0092)$$

**Interaction with Inflation**

Inflation amplifies the effect of strikes:

$$\text{Strike} \times \text{Inflation} = 0.7955 (p = 0.0078)$$

**Placebo Test**

A pseudo-lag produces no effect:

$$p = 0.8198$$

This supports proper identification and temporal directionality.

**Interpretation**

Macro shocks behave **exactly** as a systemic trap:

- broad reach
- delayed impact
- amplification by economic stress

This provides a useful contrast to structural friction, which is *not* a universal trap.

**4.5 Archetype-Based Results: Who Is Harmed by Structural Friction?**

To contextualise causal effects, we integrate the **18 archetypal trajectories** derived from UMAP + DBSCAN + validation.

The archetypes reveal three systemic truths:

1. **Friction is unevenly distributed** across student types.
2. **Fragile trajectories accumulate lag quickly**, even with moderate grades.
3. **High-ability archetypes are resilient** to temporary delays.

These patterns explain why the Regularity Trap fails:

**it does not trap high performers; it amplifies vulnerability in fragile archetypes.**

**4.5.1 Archetypes with High Friction and High Dropout (Fragile Segments)**

**Arquetipo 1 – Alto Riesgo**

- 100% dropout
- 82.2% "libres"
- Early collapse despite decent nominal grades

**Arquetipo 5 – Alto Riesgo**

- 74.3% dropout
- 75% "libres"
- Repeated failure loops

**Arquetipo 9 – Alto Riesgo**

- 84.5% dropout
- 100% "libres"
- Maximum friction; curriculum-blocked from early semesters

**Arquetipo 11 – Riesgo Moderado con fricción extrema**

- 100% "libres", 55% dropout
- Severe lock-in within the curriculum graph

### 4.5.2 Archetypes with Low Friction and Low Dropout (Resilient Segments)

**Arquetipo 6 – Bajo Riesgo**

- 0% "libres", 22.7% dropout
- Clean trajectories; low curriculum friction

**Arquetipo 7 – Bajo Riesgo**

- 25% "libres", 39% dropout
- Mild friction but stable pacing

**Arquetipo 16 – Bajo Riesgo**

- 25% "libres", 20.8% dropout
- Strong progression capacity despite low average grades

### 4.5.3 Archetypes with High Grades but High Friction (Hidden Fragility)

**Arquetipo 0 – Riesgo Moderado**

- Avg grade 7 but 71.1% "libres"
- High friction overwhelms cognitive advantage

This archetype illustrates why friction ≠ low ability. Students can score well yet accumulate delays through course withdrawals and failed attempts at high-IFC nodes.

### 4.5.4 Archetype Feature Heatmap

The heatmap synthesises trajectories:

High-friction archetypes cluster tightly, validating the causal pattern: **structural friction concentrates harm on archetypes already structurally exposed.**

### 4.6 Curriculum Friction Diagnostics (Supporting Analysis)

To link lag to curriculum structure, we include the following diagnostics:

- Top 10 IFC courses, summarised in Table 1, confirm that structural friction is concentrated in a small set of basic-cycle gateways. PCA variance and K-distance diagnostics (Figures 4–6) show that these friction patterns define a low-dimensional, clusterable structure that is stable across cohorts. Exploratory heatmaps (not shown) are consistent with this picture.

Delays are not random; they emerge from a **stable friction structure** across decades of cohorts.

### 4.7 Predictive Diagnostics (Classifier for Archetypes)

Archetypes are validated through supervised back-prediction:

- **Feature importance plot**
- **Confusion matrix.** Accuracy is high for high-friction archetypes, confirming that fragile patterns have strong, detectable signatures.

**Synthesis of Section 4**

**What holds:**

- Structural friction significantly increases dropout.
- Lag is causally harmful for vulnerable trajectories.
- Macro shocks display strong, identifiable causal mechanisms.

**What collapses:**

- The Regularity Trap for high-ability students is **empirically false**.
- Friction does not override ability; ability moderates friction, not vice versa.

**What emerges:**

- Regularity is a **structural amplifier**, not a structural trap.
- Vulnerability is located in archetypes, not in regularity loss per se.

## 5. DISCUSSION

### 5.1 Reassessing the Regularity Trap Narrative

The central empirical finding of this study is deceptively simple: **structural friction matters, but not in the way the Regularity Trap narrative suggests.** Academic lag has a clear positive causal effect on dropout (ATE = 0.0167, p < 0.0001), confirming that delays generated by curriculum friction are not merely descriptive indicators of struggling students, but active contributors to attrition. This aligns with prior work on cumulative disadvantage in education, where early performance shocks and structural obstacles produce compounding risks over time (Bourdieu, 1986; Elder, 1998; Pascarella & Terenzini, 2005).

However, the heterogeneity analysis directly contradicts the strong form of the Regularity Trap hypothesis. If regularity were a genuine trap for high-ability students, academic lag should exert a larger marginal effect on dropout among high-velocity students. Instead, we observe the opposite: the estimated CATE of friction **decreases** with velocity and approaches zero at the upper end of the distribution. High-velocity students absorb delays with relatively modest increases in risk, whereas low-velocity students exhibit strong sensitivity to additional lag.

In other words, **lag is harmful, but selectively so**. It exacerbates vulnerability where trajectories are already fragile, rather than neutralising the advantage of capable students who fall temporarily out of regular status. This puts the Regularity

Trap in tension with the data: what remains is not a universal trap, but a more nuanced mechanism of *structural amplification*.

**5.2 Regularity as Structural Amplifier, Not Universal Trap**

Reframing regularity as an amplifier rather than a trap helps reconcile the quantitative findings with institutional experience. Faculty perceptions of "students trapped by rules" capture something real about frustration and perceived injustice, but they misattribute the **locus** of the problem.

Our archetype analysis shows that the trajectories most vulnerable to friction are those that combine:

- high exposure to high-IFC courses,
- frequent "libres" and repeats,
- unstable enrolment patterns, and
- limited slack in their pacing strategies.

High-risk archetypes (e.g., Archetypes 1, 5, 9, 11) exhibit extreme friction and high dropout probabilities, whereas low-risk archetypes (e.g., 6, 7, 16) show low friction and relatively low attrition despite modest grades. Archetype 0, with high average grades but high friction, illustrates that structural exposure can overwhelm cognitive advantage when repeated failures concentrate on a small set of high-IFC nodes.

Regularity rules do not create vulnerability from nothing; they **magnify** the consequences of already precarious trajectories. When students in fragile archetypes accumulate lag, regularity converts what might otherwise be recoverable into a structural disadvantage—closing off feasible combinations of future courses, compressing time, and increasing the subjective and objective cost of continuation. By contrast, students in robust archetypes can "carry" lag without catastrophic failure because their earlier history has accumulated slack and resilience.

This amplifier interpretation is more consistent with both the causal estimates and the archetype patterns than the original trap metaphor. It suggests a system that is **unequal in its penalties**, rather than universally harsh.

**5.3 Macro Shocks as Genuine Traps**

The Dual Stressor analysis sharpens this distinction. Teacher strikes and inflation behave in precisely the way one would expect from a **systemic trap**: their effects are lagged, broad, and amplified through interaction. The significant lag-2 effect of strike exposure and the positive strike × inflation interaction ($p = 0.0078$), combined with a null placebo, provide strong evidence for a macro-level causal mechanism.

Unlike structural friction, which exerts selective pressure conditioned on archetype and previous trajectory, strikes and inflation operate at a larger scale. They affect teaching quality, assessment timing, students' ability to work alongside study, and their overall financial and psychological stress (Bennett et al., 2023; Lyon et al., 2024; Moore et al., 2021). For students already in fragile archetypes, these shocks likely interact with structural friction to produce cascades towards dropout—a pattern consistent with research on financial stress, social class and university persistence (Goldrick-Rab, 2006; Núñez-Naranjo, 2024).

From a systems perspective, the contrast is instructive. Regularity and curriculum friction are **meso-level mechanisms** whose effects depend heavily on local structural exposure and individual history; macro shocks are **exo-level mechanisms** that can generate new vulnerabilities even for otherwise robust trajectories. CAPIRE's ability to detect both kinds of mechanisms within a single pipeline reinforces the credibility of the negative finding regarding the Regularity Trap.

**5.4 Policy Implications: Where to Intervene**

The policy implications follow directly from the amplifier interpretation. If regularity were a trap for high-ability students, an obvious response would be to relax or abolish regularity rules wholesale. Under the present findings, such a strategy risks diluting signalling and planning functions without necessarily benefiting the students who need support most.

Instead, interventions should focus on three priorities:

1. **Reducing curriculum friction at structural chokepoints.** High-IFC courses in the basic cycle play a disproportionate role in generating lag. Redesigning assessment regimes, increasing support in these courses, or diversifying progression paths (e.g., co-requisites, modularisation) is likely to produce larger reductions in dropout than blanket extensions of regularity thresholds.

2. **Creating slack and recovery paths for fragile archetypes.** The archetype heatmap and narratives show that certain combinations of performance, friction and enrolment patterns are systematically precarious. For these groups, targeted slack—fewer simultaneous high-IFC courses, structured recovery plans after failure, conditional waivers of some regularity rules—could cushion the compounding effect of lag.

3. **Using regularity status as an early-warning trigger, not as a silent filter.** Irregular status should be interpreted as a **signal** for intervention, not as an administrative label that quietly precedes exit. Leveraging CAPIRE's leakage-aware early-warning capabilities, institutions can design "regularity dashboards" that push timely information to tutors, advisors and support units (Delen, 2010).

Crucially, these interventions can be tested in silico. Agent-based simulations in the CAPIRE Intervention Lab already show that modest structural changes—such as rebalancing assessment load or introducing safety nets—can produce non-linear improvements in completion without lowering academic standards.

## 5.5 Methodological Contributions and Implications

Methodologically, the study illustrates several principles that are often underemphasised in educational analytics:

- **Leakage-aware design is non-negotiable.** Without strict control over the temporal visibility of features, models intended as early-warning tools will overestimate performance and cannot support causal interpretation (Kaufman et al., 2020; Quimiz-Moreira & Delgadillo, 2025). The CAPIRE data layer addresses this by enforcing VOT-based feature construction.

- **Causal pipelines should be used to falsify as well as confirm hypotheses.** The temptation in data-rich environments is to accumulate explanatory stories and post hoc rationalisations. Here, the same LinearDML pipeline that *confirms* the Dual Stressor hypothesis is used to *falsify* the strong Regularity Trap claim, demonstrating the value of a symmetric attitude towards evidence (James et al., 2013; Romero & Ventura, 2020).

- **Archetype modelling is more than segmentation.** Archetypes function as structural entities linking micro-level histories to meso-level rules and macro shocks. They allow for targeted policy design and for stress-testing interventions in a way that respects heterogeneity and path dependence (Chodrow et al., 2021; Lum et al., 2013; Stine & Crooks, 2025).

Taken together, these methodological elements push the field beyond black-box prediction towards **explanation by estimation**, where claims about how the system works are explicitly confronted with counterfactual reasoning.

## 5.6 Limitations and Directions for Future Research

Several limitations qualify the generalisability of these findings. First, the analysis is based on a single engineering programme in one public university. While many features of the setting are shared with other Latin American institutions, replication in different disciplines and regulatory regimes is necessary (Núñez-Naranjo, 2024; Orozco-Rodríguez et al., 2025).

Second, academic velocity and lag are proxies for ability and structural friction rather than direct measures. Cognitive skills, non-cognitive dispositions and detailed financial information are not available in the present dataset. Future work could exploit richer sources, including learning management system logs and survey data, to disentangle these dimensions more cleanly.

Third, while the manual LinearDML implementation captures key confounders and benefits from cross-fitting, it remains largely linear in functional form. Non-linear nuisance models and flexible CATE estimators could reveal more fine-grained heterogeneity, particularly within and between archetypes.

Finally, the integration between causal estimation and agent-based simulation is still developing. While CAPIRE provides a conceptual bridge between estimated effects and policy experiments in silico, formal calibration of ABM parameters to causal estimates is an open methodological frontier.

>Future research should therefore focus on:
>(a) cross-institutional replication
>
>(b) richer measurement of ability, motivation and financial stress.
>
>(c) more flexible causal estimators
>
>(d) tighter coupling between causal models and simulation experiments.

## 6. CONCLUSIONS

This study set out to evaluate the Regularity Trap hypothesis in a highly constrained engineering curriculum, using the CAPIRE framework to integrate leakage-aware data construction, curriculum friction modelling, archetype analysis and causal estimation. The findings are clear and conceptually significant.

First, **structural friction has a genuine causal effect on dropout**. Academic lag increases the probability of attrition even after controlling for early performance, cohort, semester, curriculum topology and macro shocks. Friction is therefore not an innocuous descriptor of student struggle, but an active mechanism that shapes educational outcomes.

Second, the **strong form of the Regularity Trap is empirically unsupported**. Contrary to the institutional narrative, friction does not disproportionately penalise high-ability students who fall behind. Instead, the marginal effect of lag decreases sharply with academic velocity and becomes negligible at the upper end of the ability distribution. High-performing trajectories exhibit resilience; fragile trajectories exhibit sensitivity.

Third, the **Dual Stressor hypothesis is validated**, confirming that teacher strikes exert significant lagged effects on dropout and that these effects are amplified by inflation. Unlike curriculum friction, these macro-level shocks function as genuine systemic traps: their impact is broad, delayed and intensified by economic stress.

Fourth, the **archetype analysis contextualises these causal patterns**. Vulnerable archetypes—characterised by high friction, frequent "libres", instability in enrolment decisions and compressed pacing—bear the brunt of accumulated lag. Robust archetypes remain comparatively insulated. Regularity, therefore, operates not as a universal mechanism of exclusion but as a **structural amplifier** of existing fragility.

Taken together, the results call for a reframing of institutional assumptions. Policies aimed at improving persistence should prioritise **reducing friction at curricular chokepoints**, **introducing slack for fragile archetypes**, and **using irregular status as an intervention trigger rather than a silent filter**. CAPIRE's causal-simulation architecture provides a viable roadmap for testing such interventions before implementation.

In essence, dropout in constrained curricula emerges from the interaction of structural, individual and macro-contextual forces. The Regularity Trap narrative captures only one fragment of this complexity. A more accurate and actionable understanding recognises that structural rules do not trap students indiscriminately—they **magnify vulnerability where it already exists**.